\documentclass[11pt]{article}

\pdfoutput=1

\usepackage[T1]{fontenc}
\usepackage[latin9]{inputenc}
\usepackage[a4paper]{geometry}
\usepackage[active]{srcltx}
\usepackage{amsmath}
\usepackage{amssymb}
\usepackage{esint}
\usepackage{ulem}
\usepackage{mathrsfs}

\usepackage{xcolor}

\makeatletter
\usepackage{textcomp}

\pdfoutput=1 % if your are submitting a pdflatex (i.e. if you have
\usepackage{jheppub}% for details on the use of the package, please
\usepackage{etoolbox}% http://ctan.org/pkg/etoolbox
    
    \patchcmd{\maketitle}{\@fpheader}{}{}{}

\newcommand*\xbar[1]{%
  \hbox{%
    \vbox{%
      \hrule height 0.5pt % The actual bar
      \kern0.3ex%         % Distance between bar and symbol
      \hbox{%
        \kern-0.0em%      % Shortening on the left side
        \ensuremath{#1}%
        \kern-0.0em%      % Shortening on the right side
      }%
    }%
  }%
}

\usepackage{amsfonts}

\usepackage{bbm}

\newcommand{\be}{\begin{equation}}
\newcommand{\ee}{\end{equation}}
\newcommand{\bea}{\begin{eqnarray}}
\newcommand{\eea}{\end{eqnarray}}

\title{\boldmath  Asymptotic $\mathcal O(r)$ gauge symmetries and gauge-invariant Poincar\'e generators in higher spacetime dimensions}

\author{Oscar Fuentealba}

\affiliation{Universit\'e Libre de Bruxelles and International Solvay Institutes, ULB-Campus Plaine CP231, B-1050 Brussels, Belgium}

\emailAdd{oscar.fuentealba@ulb.be}

\preprint{}

\abstract{The asymptotic symmetries of electromagnetism in all higher spacetime dimensions $d>4$ are extended, by incorporating consistently angle-dependent $u(1)$ gauge transformations with a linear growth in the radial coordinate at spatial infinity. Finiteness of the symplectic structure and preservation of the asymptotic conditions require to impose a set of strict parity conditions, under the antipodal map of the $(d-2)$-sphere, on the leading order fields at infinity. Canonical generators of the asymptotic symmetries are obtained through standard Hamiltonian methods. Remarkably, the theory endowed with this set of asymptotic conditions turns out to be invariant under a six-fold set of angle-dependent $u(1)$ transformations, whose generators form a centrally extended abelian algebra. The new charges generated by the $\mathcal O(r)$ gauge parameter are found to be conjugate to those associated to the now improper subleading $O(r^{-d+3})$ transformations, while the standard $\mathcal O(1)$ gauge transformations are canonically conjugate to the  subleading $\mathcal{O}(r^{-d+4})$ transformations. This algebraic structure, characterized by the  presence of central charges, allows us to perform a nonlinear redefinition of the Poincar\'e generators, that results in the decoupling of all of the $u(1)$ charges from the Poincar\'e algebra. Thus, the mechanism previously used in $d=4$ to find gauge-invariant Poincar\'e generators is shown to be a robust property of electromagnetism in all spacetime dimensions $d\geq 4$.}

\makeatother

\begin{document}
\maketitle \flushbottom

\section{Introduction}

In light of the very recent results for gravity and electromagnetism \cite{Fuentealba:2022xsz,Fuentealba:2023rvf}, where it was possible to extend the asymptotic symmetries of the theories by consistently accommodating logarithmic branches and a linear $u(1)$ symmetry (namely, generated by an $\mathcal O(r)$ gauge parameter) at spatial infinity, we address the direct question about whether similar extensions can be found in the case of electromagnetism in higher spacetime dimensions $d>4$.

In the case of four-dimensional gravity in \cite{Fuentealba:2022xsz}, by following the approach in \cite{Henneaux:2018cst,Henneaux:2018hdj,Henneaux:2019yax}, a logarithmic behaviour in the fall-off of the metric was included. This enabled us to extend the renowned standard BMS algebra \cite{Bondi:1962px,Sachs:1962wk,Sachs:1962zza} by a new family of angle-dependent logarithmic supertranslations. The latter together with standard supertranslations, were shown to satisfy an abelian algebra with non-trivial central charges. Now, it is precisely the presence of these central charges which allowed, through a suitable nonlinear redefinition of the Lorentz generators, to disentangle the whole set of pure supertranslations (standard and logarithmic ones) from the Lorentz algebra, finding in this manner a resolution to the angular momentum ambiguity under supertranslations at spatial infinity. Note that different mechanisms have been developed at null infinity \cite{Mirbabayi:2016axw,Bousso:2017dny,Javadinezhad:2018urv,Javadinezhad:2022hhl} (see also \cite{Chen:2021szm,Chen:2021zmu,Compere:2023qoa} and references therein).

Analogue enhancements were explored in electromagnetism. Concretely, a relaxation of the asymptotic conditions of \cite{Henneaux:2018gfi} was performed in \cite{Fuentealba:2023rvf}. As a result of this analysis, it was consistently  accommodated not only a logarithmic but a linear growth (associated to the subleading soft photon theorems \cite{Lysov:2014csa,Campiglia:2016hvg,Conde:2016csj}) for the gauge potential at infinity. As in the case of gravity, central charges emerged between different types of angle-depedent gauge transformations, to wit, linear, logarithmic, standard and subleading $u(1)$ transformations. It was then possible to redefine the Poincar\'e generators, in such a way that the asymptotic symmetry algebra becomes the direct sum of the Poincar\'e algebra and a centrally extended abelian algebra associated to the $u(1)$ symmetries.

Then, it is natural to wonder whether a similar extension of the asymptotic symmetries  can be performed in higher dimensions. Spatial infinity yields the perfect arena to address this question, since one can avoid the complications associated to the fractional decay in the radial coordinate of the radiative branch of the fields (in odd dimensions), which could lead to problems with the conformal definition of null infinity \cite{Hollands:2003ie,Hollands:2004ac,Tanabe:2011es}. Indeed, in the exhaustive study by Henneaux and Troessaert in \cite{Henneaux:2019yqq}, it was shown how to extend the Hamiltonian formulation of electromagnetism in dimensions $d>4$, such that a two-fold family of angle-dependent $u(1)$ transformations emerges at spatial infinity (see also \cite{Esmaeili:2019hom}, where stronger conditions are imposed, such that only one angle-dependent $u(1)$ symmetry is found). Unlike the $d=4$ case, parity conditions on the fields are not needed (being not inconsistent to impose them though, according to \cite{Henneaux:2019yqq}). Indeed, they explicitly integrated the equations of motion at spatial infinity and then performed the match with null infinity, which led to non divergent terms in the asymptotic behaviour of the fields without imposing any type of parity conditions. This latter fact enabled them to find generalizations of the matching conditions between the fields at the past of future null infinity $\mathscr I^{+}_{-}$ and the future of past null infinity $\mathscr I^{-}_{+}$.

\vskip 2mm

Here, we report a set of boundary conditions for electromagnetism in higher dimensions that are invariant under an improper (or large) $u(1)$ gauge parameter growing linearly at spatial infinity (by ``improper'' we refer to an asymptotic gauge symmetry with a non vanishing charge, according to the terminology introduced in \cite{Benguria:1976in}). This is achieved by imposing parity conditions on the asymptotic behaviour of the fields under the antipodal map of the $(d-2)$-sphere. We have found that the asymptotic symmetries are given by a six-fold family of angle-dependent $u(1)$ transformations, whose canonical generators satisfy an abelian algebra endowed with non-zero central charges. 

\vskip 2mm

We recall that in the analysis of \cite{Henneaux:2019yqq}, absence of magnetic sources was assumed in order to tame the Lorentz boosts. However, we shall show that this last assumption can be relaxed by taking a different boundary contribution to the symplectic form, which is necessary for the further extension that allows the linear growth in the gauge potential. In fact, in the absence of the linear behaviour, this reformulation of the theory leads to a four-fold family of angle-dependent $u(1)$ transformations, since subleading transformations generated by a parameter with decay $\sim r^{-d+4}$ become improper (in contrast with the two angle-dependent $u(1)$ symmetries found in \cite{Henneaux:2019yqq}). The corresponding canonical generators are shown to form a centrally extended abelian algebra.  It must be mentioned that the role of the logarithmic gauge transformations in $d=4$ found in \cite{Fuentealba:2023rvf} is now played by the improper subleading $\mathcal O(r^{-d+4})$ gauge transformations in all higher dimensions $d>4$ (note that these subleading $u(1)$ symmetries turn out to be analogue to the subleading supertranslations in higher dimensions found in \cite{Fuentealba:2021yvo,Fuentealba:2022yqt}). 

Consistency does not require any parity condition so far. However, once the linear relaxation of the gauge potential is included, finiteness of the symplectic structure requires to impose parity conditions on the leading order fields. In this formulation (with the linear growth), the gauge parameter with subleading decay $\sim r^{-d+3}$ possesses a nonvanishing charge. Hence, the system is invariant under a six-fold family of angle-dependent $u(1)$ transformations. The algebra of the asymptotic symmetries acquires an additional nonvanishing central charge along the bracket of the new linear $u(1)$ charges and those generated by subleading $O(r^{-d+3})$ gauge transformations. Note that the required parity conditions do not discard any known solution of the theory. 
 
Our generalized set of boundary conditions enables us to apply the mechanism that decouples all of the improper $u(1)$ gauge transformations (linear, standard and the subleading ones) from the Poincar\'e algebra (given the presence of invertible central charges among the $u(1)$ canonical generators). This explicitly shows that this decoupling, which transforms the asymptotic symmetry algebra in the direct sum form, is a general property of electromagnetism in all higher spacetime dimensions rather than an accident in $d=4$.

\vskip 2mm

The plan of the paper goes as follows. In Section \ref{sec:Section2}, the extended Hamiltonian formulation for electromagnetism in higher dimensions is presented. This comprises the asymptotic behaviour and parity conditions of the fields that ensure a well-defined action of the Poincar\'e symmetry. Finiteness of the symplectic structure and equations of motion are also discussed. We are then on solid grounds to apply standard Hamiltonian methods for the computation of the canonical generators of the gauge symmetries, which lead to finite and integrable charges, an analysis that is performed in Section \ref{sec:Section3}. The canonical realization of the Poincar\'e symmetry is carried out in Section \ref{sec:Section4}, where we obtain that the asymptotic symmetry algebra is given by the semi-direct sum of the Poincar\'e algebra and a centrally extended abelian algebra. In Section  \ref{sec:Section5}, we decouple all of the gauge transformations from the Poincar\'e algebra through an appropriate redefinition of the Poincar\'e generators. Section \ref{sec:Section6} is devoted to some concluding remarks. In Appendix \ref{app:AppendixA}, we construct the symplectic form of the theory and prove Poincar\'e invariance. Finally, in Appendix \ref{app:AppendixB}, we give  the complete list of the transformation laws of the fields obtained from the preservation of the asymptotic conditions under the Poincar\'e symmetry.

\section{Action principle and boundary conditions}\label{sec:Section2}

In order to accomplish a consistent relaxation of the boundary conditions that are invariant under angle-dependent $u(1)$ transformations, we consider from  the beginning the extended Hamiltonian formulation of the theory. Indeed, this is the action that one obtains after direct application of the Dirac formalism for systems with gauge symmetries \cite{Dirac1967,Henneaux:1992ig}. This has been shown to be useful in previous treatments of the problem in the context of electromagnetism, see e.g. \cite{Henneaux:2018gfi,Henneaux:2019yqq,Fuentealba:2023rvf}. The extended Hamiltonian action principle in $d$ spacetime dimensions on a Minkowski background reads
\begin{equation}
I_{H}[A_{i},\pi^{i},A_{0},\pi^{0};\psi,\lambda]= \int dt  \int d^{d-1}x\Big[\pi^{i}\dot{A}_{i}+ \pi^0\dot{A}_0- (\mathcal{H}+\tilde \epsilon \mathcal G + \tilde \mu \pi^0)  - \psi \mathcal{G}- \lambda \pi^0 \Big]  +B_{\infty} \, , \label{eq:Action-Extended0} \\
\end{equation}
where
\begin{equation}\label{eq:Ham-density}
\mathcal H = \frac{1}{2\sqrt{g}}\pi^{i}\pi_{i} +\frac{\sqrt{g}}{4}F^{ij}F_{ij}  +A_0 \mathcal{G}-\partial^i\pi^0 A_i \,.
\end{equation}
The phase space is described by the conjugate canonical pairs $(A_i,\pi^i)$ and $(A_0,\pi^0)$. The surface term at spatial infinity $B_{\infty}$ depends on the boundary conditions and its precise form will be shown below (once we provide our set of asymptotic conditions). Variation with respect to the Lagrange multipliers $\psi$ and $\lambda$ enforces the constraints
\begin{equation}
\mathcal G=-\partial_i\pi^i=0\,,\qquad \pi^0=0\,,
\end{equation}
respectively. The theory is thus invariant under the following gauge transformations
\begin{equation}
\delta_\epsilon A_i=\partial_i \epsilon\,,\qquad \delta_i A_0=\mu\,.
\end{equation}
The vector field that generates Poincar\'e transformations is
\begin{eqnarray}
\xi&=& b_i x^i+a^{\perp}\,,\\
\xi^i&=& b^i_{\,\,j}x^j+a^i\,.
\end{eqnarray}
The arbitrary constants $b_i$, $b_{ij}=-b_{ji}$, $a^{\perp}$ and $a^{i}$ parametrize Lorentz boosts, spatial rotations and standard translations, respectively. Note that the term $-b^ix^0$ that comes from $\xi^i$ can be eliminated by a spatial translation (at any given time). 

The infinitesimal transformation laws of the fields under Poincar\'e deformations \cite{Henneaux:2019yqq}  (see also \cite{Henneaux:2018gfi}) are given by 
\begin{align}
\delta_{\xi,\xi^{i}}A_{i} & =\frac{\xi\pi_{i}}{\sqrt{g}}+\partial_{i}\left(\xi A_0 \right)+\mathcal{L}_{\xi}A_{i}+\partial_i \epsilon_{(\xi,\xi^i)}\,, \label{eq:A_i}\\
\delta_{\xi,\xi^{i}}\pi^{i} & =\sqrt{g}\nabla_{m}\big(F^{mi}\xi\big)+\xi\partial^{i}\pi^0+\mathcal{L}_{\xi}\pi^{i}\,,\label{eq:pi_i}\\
\delta_{\xi,\xi^{i}}A_0 & =\nabla_{i}\left(\xi A^{i}\right)+\xi^i \partial_i A_0+\mu_{(\xi,\xi^i)}\,,\label{eq:A_0}\\
\delta_{\xi,\xi^{i}}\pi^0 & =\xi\partial_{i}\pi^{i}+\partial_i(\xi^i\pi^0)\,, \label{eq:pi_0}
\end{align}
where the spatial Lie derivatives on the fields along $\xi^i$ read
\begin{align}
\mathcal L_{\xi}A_i &=\xi^j \partial_j A_i+\partial_i \xi^j A_j\,,\\
\mathcal L_{\xi}\pi^i &=\partial_j(\xi^j\pi^i)-\partial_j \xi^i \pi^j\,.
\end{align}
Note that the action of the Poincar\'e group is canonical only if we apply the additional corrective gauge transformations generated by $\epsilon_{(\xi,\xi^i)}$ and $\mu_{(\xi,\xi^i)}$. The precise form of these parameters will be shown when we discuss the canonical realization of the Poincar\'e symmetry in Section \ref{sec:Section4}.

Consistency of the formulation for the boundary conditions imposed here, requires that Hamiltonian density in \eqref{eq:Action-Extended0} is given by the sum of the usual term $\mathcal H$ in \eqref{eq:Ham-density}  and additional terms proportional to the constraints $\mathcal G$ and $\pi^0$, where
\begin{align}
\tilde{\epsilon} & =\frac{1}{r^{d-4}}\tilde{\epsilon}^{(d-4)}+\frac{1}{r^{d-3}}\tilde{\epsilon}^{(d-3)}+\mathcal{O}\left(r^{-d+2}\right)\,,\\
\tilde{\mu} & =\frac{1}{r^{d-3}}\tilde{\mu}^{(d-4)}+\frac{1}{r^{d-2}}\tilde{\mu}^{(d-3)}+\mathcal{O}\left(r^{-d+1}\right)\,,
\end{align}
with
\begin{align}
\tilde{\epsilon}^{(d-4)}&=-\Psi^{(d-5)}\,,\qquad \tilde{\epsilon}^{(d-3)}=-\frac{1}{d-3}\Psi^{(d-4)}\,,\\
\tilde{\mu}^{(d-4)}&=-\xbar \triangle \Phi^{(d-5)}+2(d-5)\Phi^{(d-5)}\,,\\
\tilde{\mu}^{(d-3)}&=-\frac{1}{d-3}\Big[\xbar D_A \xbar A^A+(2d-5)\xbar A_r\Big]\,.
\end{align}
Concretely, these terms must be added due to the corrective gauge transformations necessary to have a canonical action of the Poincar\'e symmetry. The coefficients $\xbar A_A$, $\xbar A_r$, $\Phi^{(d-5)}$, $\Psi^{(d-5)}$ and $\Psi^{(d-5)}$ are arbitrary functions of the $(d-2)$-sphere at spatial infinity, which appear in the radial fall-off of the canonical fields as we shall show below.

Following the approach introduced in \cite{Henneaux:2018hdj,Henneaux:2019yax} (see also \cite{Henneaux:2018gfi,Fuentealba:2023rvf} for the Maxwell theory in $d=4)$, we will deform the usual decay of the gauge potential $(\tilde{A}_0,\tilde{A}^i)\sim (r^{-1},r^{-d+3})$ (obtained from the fall-off of the electric and magnetic fields at spatial infinity $\sim r^{-d+2}$), by  improper gauge transformations, namely,
\begin{align}
A_0 &=\Xi+\tilde{A}_0 \,, \label{eq:A0-decay}\\ 
A_i&=\partial_i \Phi +\tilde{A}_i\,, \label{eq:Ai-decay}
\end{align}
where we will impose that
\begin{eqnarray}
\Xi&=&\Psi_{\text{lin}}\,,\\
\Phi&=& r\Phi_{\text{lin}}+\xbar \Phi+\frac{1}{r}\Phi^{(1)}+\dots+\frac{1}{r^{d-5}}\Phi^{(d-5)}\,.
\end{eqnarray}
Thus, in polar coordinates this fall-off reads
\begin{eqnarray}
A_0&=&\Psi_{\text{lin}}+\frac{1}{r}\xbar \Psi+\frac{1}{r^2}\Psi^{(1)}+\mathcal O\left(r^{-3}\right)\,,\\
A_r&=&\partial_r \Phi+\frac{1}{r^{d-3}}\xbar A_r +\frac{1}{r^{d-2}}A^{(2)}_r+\mathcal O\left(r^{-d+1}\right)\,,\\
A_A&=&\partial_A \Phi+\frac{1}{r^{d-4}}\xbar A_A+\frac{1}{r^{d-3}}A^{(2)}_A+\mathcal O\left(r^{-d+2}\right)\,,
\end{eqnarray}
while the one for the conjugate momenta (of weight density one) goes as follows
\begin{align}
\pi^0 & =\frac{1}{r^{2}}\pi_{\Psi}^{(2)}+\mathcal O\left(r^{-3}\right)\,,\\ \label{eq:DecayPi0}
\pi^{r} & =\xbar\pi^{r}+\frac{1}{r}\pi_{(2)}^{r}+\frac{1}{r^{2}}\pi_{(3)}^{r}+\mathcal O\left(r^{-3}\right)\,,\\
\pi^{A} & =\frac{1}{r}\xbar\pi^{A}+\frac{1}{r^{2}}\pi_{(2)}^{A}+\frac{1}{r^{3}}\pi_{(3)}^{A}+\mathcal O\left(r^{-4}\right)\,.
\end{align}
The coefficients in the radial expansions of the fields are functions of the angles $\xbar x^A$ of the round unit $(d-2)$-sphere at spatial infinity. The indices $A,B,C,\dots$ are raised and lowered by the metric $\xbar g_{AB}$ on the $(d-2)$-sphere, where $\xbar D_A$ stands for its associated covariant derivative.

The fall-off of the Lagrange multipliers is chosen to be the same as the proper gauge transformations (which are established in Section \ref{sec:Section3}), namely,
\begin{equation}
\psi=\frac{\psi^{(1)}}{r} +\mathcal O\left(r^{-2}\right)\,,\qquad \lambda=\frac{\lambda^{(1)}}{r^2}+\mathcal O\left(r^{-3}\right)\,,
\end{equation} 
with $\psi^{(d-4)}=\psi^{(d-3)}=0$ and $\lambda^{(d-4)}=\lambda^{(d-3)}=0$.

Finiteness of the symplectic structure and preservation under Poincar\'e symmetry require that the coefficients in the asymptotic behaviour of the fields must satisfy the following parity conditions under the antipodal map of the $(d-2)$-sphere $(\xbar x^A\rightarrow -\xbar x^A)$,
\begin{align}
\xbar A_{r}&=(\xbar A_{r})_{\text{odd}}-(d-4)\Phi^{(d-4)}\,,\quad\xbar\pi^{r}=\text{even}\,,\label{eq:Parity-cond-A-core}\\
\xbar A_{A}&=(\xbar A_{A})_{\text{even}}+\partial_{A}\Phi^{(d-4)}\,,\quad\xbar\pi^{A}=\text{odd}\,,\\ \label{eq:Parity-cond-core}
\Psi_{\text{lin}}&=\text{even}\,,\quad \Phi_{\text{lin}}=\text{odd}\,,\quad \Phi^{(d-4)}=\text{even}\,.
\end{align}
Note that in addition to the strict parity conditions obeyed by the leading order fields, the ``core'' fields, i.e., leading order fields for the usual decay (already denoted by $(\tilde{A}_0,\tilde{A}^i)$ in \eqref{eq:A0-decay} and \eqref{eq:Ai-decay}, respectively), are subject to a class of twisted parity conditions as $\xbar A_r$ and $\xbar A_A$ do not have a definite parity. If we set $d=4$, $\xbar A_r$ would be purely odd, while the function $\xbar A_A$ remains as a strict parity (even) function deformed by a gauge transformation that generates a term of the opposite parity, recovering thus the parity properties of the fields in \cite{Henneaux:2018gfi}.

Our set of boundary conditions is completed by imposing the faster fall-off for the Gauss constraint $\partial_i\pi^i \sim r^{-3}$ (in polar coordinates), which implies the following asymptotic relations
\begin{equation}\label{eq:Gauss-constraint}
\partial_A \xbar \pi^A=0\,,\qquad \partial_A \pi^A_{(2)}-\pi^r_{(2)}=0\,.
\end{equation}

We are now in position to bring the specific form of the surface term at infinity $B_{\infty}$ of the action principle in \eqref{eq:Action-Extended0}. This is given by two different terms $\mathcal S$ and $\mathcal B$. The former is linear in time derivatives on the canonical variables, contributing then to the symplectic two-form $\Omega$ of the theory. The second one, $\mathcal B$, is quadratic in the canonical variables, without time derivatives on the fields. Thus, we have that
\begin{equation}
B_{\infty}=\int dt \oint_{S^{\infty}_{d-2}} d^{d-2}x \left(\mathcal S-\mathcal B\right)\,,
\end{equation}
where
\begin{eqnarray}
\mathcal S&=&-\sqrt{\xbar g}\left[\xbar A_r \dot{\xbar \Psi}+(d-4)\Psi^{(d-4)}\dot{\xbar \Phi}+A^{(2)}_r \dot{\Psi}_{\text{lin}}+(d-3)\Psi^{(d-3)}\dot{\Phi}_{\text{lin}}\right]\,,\label{eq:S}\\
\mathcal B&=&\left[\xbar\pi^{r}-(d-4)\sqrt{\xbar g}\,\Psi^{(d-4)}\right]\Psi_{\text{lin}}+\sqrt{\xbar g}\,\partial_{A}\xbar A_{r}\xbar D^{A}\Phi_{\text{lin}}-(d-2)\sqrt{\xbar g}\,\xbar A_{r}\Phi_{\text{lin}}\,.
\end{eqnarray}

Here, we must mention the difference with the boundary conditions adopted in \cite{Henneaux:2019yqq}. In absence of linear fields $(\Phi_{\text{lin}}=\Psi_{\text{lin}}=0)$, instead of asking for a faster fall-off of the magnetic field through the condition $\xbar A_A=\xbar D_A \Theta$ in order to tame the Lorentz boosts, we managed to solve it by introducing the field $\Psi^{(d-4)}$ (associated to a subleading coefficient of $A_0$ in \eqref{eq:A0}) in the boundary contribution to the kinetic term. This field is conjugate to the leading order field $\xbar \Phi$. The introduction of $\Psi^{(d-4)}$ will have a direct implication in the subleading gauge symmetry associated to the shifts in $\xbar A_r$, which becomes improper (as shown in Section \ref{sec:Section3}). Note that the boundary term $\mathcal B$ vanishes in this case. Then, if we switch on the linear growth, similar terms appear in $\mathcal S$, where the linear fields are conjugate to sub-subleading coefficients in the radial expansion of the gauge field $(A_r^{(2)},\Psi^{(d-3)})$. In this case, the boundary term $\mathcal B$ is necessary in order to have a well-defined action principle. The explicit construction of the boundary contribution $\mathcal S$ to the symplectic form is performed in Appendix \ref{app:AppendixA}.

\vskip 2mm

Given the full set of boundary conditions, we proceed to show the finiteness of the symplectic structure. Using the fall-off of the canonical fields, a rather direct calculation implies that
\begin{align}
\int dt d^{d-1}x\,\pi^{\mu}\dot{A}_{\mu}=\int dtdr\oint d^{d-2}x\Big[ & \xbar\pi^{r}\dot{\Phi}_{\text{lin}}+\xbar\pi^{A}\partial_{A}\dot{\Phi}_{\text{lin}}  \nonumber\\
&+\frac{1}{r}\left(\xbar\pi^{A}\partial_A \dot{\Phi}+\pi_{(2)}^{r}\dot{\Phi}_{\text{lin}}+\pi_{(2)}^{A}\partial_{A}\dot{\Phi}_{\text{lin}}\right)+\mathcal O\left(r^{-2}\right)\Big]\,.
\end{align}
It is clear that the linear divergence vanishes by virtue of the parity conditions as the term $\xbar\pi^{r}\dot{\Phi}_{\text{lin}}+\xbar\pi^{A}\partial_{A}\dot{\Phi}_{\text{lin}}$ is odd, and it is then zero under the integral on the $(d-2)$-sphere. The remaining terms, along the logarithmic divergence, vanish (after integration by parts) by virtue of the faster fall-off of the Gauss constraint (after use of relations \eqref{eq:Gauss-constraint}).

Note that in absence of the linear growth, parity conditions are not needed. This is in agreement with \cite{Henneaux:2019yqq}.

\vskip 4mm

We have now all the elements to analyse the equations of motion for the extended Hamiltonian formulation of the theory with this enlarged set of boundary conditions.
\begin{itemize}
\item As aforementioned, variation of the Hamiltonian action \eqref{eq:Action-Extended0} with respect to the Lagrange multipliers $\psi$ and $\lambda$ enforce the constraints $\mathcal G=-\partial_i\pi^i$ and $\pi^0=0$.
\item The variation with respect to the spatial component of the gauge potential $A_i$ gives a bulk term and a boundary term, which should vanish independently. The bulk term imposes the following equation of motion for the spatial conjugate momentum
\begin{equation}\label{eq:momentum}
\dot{\pi}^{i}=-\sqrt{g}\nabla_i F^{ij}+\nabla^i\pi^0\,.
\end{equation}
Vanishing of the boundary term in turn implies the following dynamical equations
\begin{align}
\dot{\Psi}_{\text{lin}}&=0\,,\qquad\dot{\Psi}^{(d-4)}=0\,, \label{eq:A01}\\
\dot{\xbar \Psi}&=\xbar \triangle\,\Phi_{\text{lin}}+(d-2)\Phi_{\text{lin}}\,,\label{eq:A02}\\
\dot{\Psi}^{(d-3)}&=-\frac{1}{d-3}\left[(d-2)\xbar A_r-(d-4)\xbar D_A\xbar A^A\right]\,. \label{eq:A03}
\end{align}
If we now expand asymptotically the equation of motion for the spatial component of the conjugate momentum in \eqref{eq:momentum}, we obtain that
\begin{equation}
\dot{\xbar \pi}^r=0\,,\qquad \dot{\pi}^r_{(2)}=\sqrt{\xbar g}\left[\xbar \triangle\,\xbar A_r+(d-4)\xbar D_A \xbar A^A\right]\,.
\end{equation}
From here we can see that
\begin{eqnarray}
\partial_t\left[\xbar \pi^r-(d-4)\sqrt{\xbar g}\Psi^{(d-4)}\right]&=&0\,,\\
\partial_t\left[\pi^r_{(2)}-(d-3)\sqrt{\xbar g}\Psi^{(d-3)}\right]&=&\sqrt{\xbar g}\left[\xbar \triangle\,\xbar A_r+(d-2)\xbar A_r\right]\,.\label{eq:dipole-moment}
\end{eqnarray}
\item Variation of the action with respect to $\pi^i$ enforces the following equation of motion of the gauge potential
\begin{equation}\label{eq:dot-potential}
\dot{A}_i=\frac{\pi_i}{\sqrt{\xbar g}}+\partial_i\left(A_0+\psi+\tilde{\epsilon}\right)\,,
\end{equation} 
where the asymptotic expansion of its radial component implies that
\begin{align}
\dot{\Phi}_{\text{lin}}&=0\,,\qquad\xbar A_r=0\,, \label{eq:E1}\\
\dot{A}^{(2)}_r&=\frac{1}{\sqrt{\xbar g}}\left[\xbar \pi^r-(d-4)\sqrt{\xbar g}\Psi^{(d-4)}\right]\,. \label{eq:E2}
\end{align}
The asymptotic expansion of the angular component of \eqref{eq:dot-potential} gives the conditions $\partial_A \dot{\Phi}_{\text{lin}}=0$ (which is immediately satisfied) and
\begin{equation}
\partial_A\left(\dot{\xbar \Phi}-\Psi_{\text{lin}}\right)=0\,. \label{eq:E3}
\end{equation}
\item The variation with respect to the time component of the gauge potential $A_0$, gives also a bulk a term and a boundary term. However, the bulk term vanishes by virtue of the constraints $\mathcal G=-\partial_i\pi^i$ and $\pi^0=0$, while the boundary term is zero provided the following equations are satisfied
\begin{align}
\dot{\Phi}_{\text{lin}}&=0\,,\qquad \dot{\xbar A}_r=0\,,\\
\dot{\xbar \Phi}&=\Psi_{\text{lin}}\,,\label{eq:Phi-dot}\\
\dot{A}^{(2)}_r&=\frac{1}{\sqrt{\xbar g}}\left[\xbar \pi^r-(d-4)\sqrt{\xbar g}\Psi^{(d-4)}\right]\,.\label{eq:A2r-dot}
\end{align}
These are in agreement with the asymptotic expansion of the equations of motion obtained from the variation with respect to the spatial conjugate momentum in \eqref{eq:E1}, \eqref{eq:E2} and \eqref{eq:E3}. 
\item Finally, we take the variation of the action with respect to $\pi^0$. This yields the following equation of motion for $A_0$
\begin{equation}
\dot{A}_0=\partial_i A^i+\tilde{\mu}+\lambda\,.
\end{equation}
Note that the radial expansion of the above equations leads to the equations of motion for the boundary fields obtained in \eqref{eq:A01}, \eqref{eq:A02} and \eqref{eq:A03}. 
\end{itemize}

Thus, we have explicitly shown that our asymptotic conditions provide a true extremum of the extended Hamiltonian action, where the dynamical equations obtained from the boundary terms are fully compatible with the equations of the motion for the canonical variables.

\section{A six-fold set of angle-depedent $u(1)$ gauge transformations}\label{sec:Section3}

This new set of asymptotic conditions turns out to be preserved by gauge parameters with a relaxed fall-off, compared with \cite{Henneaux:2019yqq}, that is to say
\begin{align}
\epsilon & =r\epsilon_{\text{lin}}+\xbar\epsilon+\frac{1}{r}\epsilon^{(1)}+\dots+\frac{1}{r^{d-4}}\epsilon^{(d-4)}+\frac{1}{r^{d-3}}\epsilon^{(d-3)}+\mathcal{O}\left(r^{-d+2}\right)\,,\\
\mu & =\mu_{\text{lin}}+\frac{1}{r}\xbar\mu+\frac{1}{r^{2}}\mu^{(1)}+\dots+\frac{1}{r^{d-3}}\mu^{(d-4)}+\frac{1}{r^{d-2}}\mu^{(d-3)}+\mathcal{O}\left(r^{-d+1}\right)\,.
\end{align}
where the parity conditions of the leading order fields are given by
\begin{equation}
\epsilon_{\text{lin}}=\text{odd}\,,\quad\mu_{\text{lin}}=\text{even}\,.
\end{equation}
The remaining functions in the radial expansions of the parameters are no subject to parity conditions in principle. However, as we will see only the even component of $\epsilon^{(d-3)}$ and the odd component of $\mu^{(d-3)}$ generate improper gauge transformations.
The subleading functions $\epsilon^{(d-4)}$ and  $\mu^{(d-4)}$ (being completely arbitrary) also generate improper gauge transformations (see below).

The boundary conditions are preserved provided the following transformation laws are satisfied
\begin{equation}
\delta_{\epsilon}\Phi_{\text{lin}}=\epsilon_{\text{lin}}\,,\quad\delta_{\epsilon}\xbar \Phi=\xbar\epsilon\,,\quad\delta_{\epsilon}\xbar A_{r}=-(d-4)\epsilon^{(d-4)}\,,\quad\delta_{\epsilon}A_{r}^{(2)}=-(d-3)\epsilon^{(d-3)}\,,
\end{equation}
\begin{equation}
\delta_{\epsilon}\xbar A_{A}=\partial_A \epsilon^{(d-4)}\,,\quad\delta_{\epsilon}A_{A}^{(2)}=\partial_A \epsilon^{(d-3)}\,,
\end{equation}
\begin{equation}
\delta_{\mu}\Psi_{\text{lin}}=\mu_{\text{lin}}\,,\quad\delta_{\mu}\xbar\Psi=\xbar\mu\,,\quad\delta_{\mu}\Psi^{(d-4)}=\mu^{(d-4)}\,,\quad\delta_{\mu}\Psi^{(d-3)}=\mu^{(d-3)}\,.
\end{equation}

In order to compute the canonical generators $G_{\epsilon,\mu}$ of the gauge symmetries, we will make use of the (non-degenerate) symplectic form of the theory
\begin{align}
\Omega & =\int d^{d-1}x\big(d_{V}\pi^{i}d_{V}A_{i}+d_{V}\pi^0 d_{V}A_0\big)\nonumber\\
 & \quad-\oint d^{d-2}x\sqrt{\xbar g}\Big[d_{V}\xbar A_{r}d_{V}\xbar\Psi+(d-4)d_{V}\Psi^{(d-4)}d_{V}\xbar \Phi \nonumber\\
 &\qquad \qquad \qquad \qquad+d_{V}A_{r}^{(2)}d_{V}\Psi_{\text{lin}}+(d-3)d_{V}\Psi^{(d-3)}d_{V}\Phi_{\text{lin}}\Big]\,,\label{eq:Omega0}
\end{align}
through the formula $\iota_{X}\Omega=-d_V G_X$ (where $X$ is a Hamiltonian vector field generating the phase space transformations). Thus, we obtain that
\begin{equation}
G_{\epsilon,\mu}=\int d^{d-1}x\left(\mu \pi^0-\epsilon\partial_i \pi^i\right)+Q_{\epsilon,\mu}\,,
\end{equation}
where the surface term is given by
\begin{equation}
Q_{\epsilon,\mu}=Q_{\epsilon_{\text{lin}}}+Q_{\xbar \epsilon}+Q_{\epsilon^{(d-4)}}+Q_{\epsilon^{(d-3)}}+Q_{\mu_{\text{lin}}}+Q_{\xbar \mu}+Q_{\mu^{(d-4)}}+Q_{\mu^{(d-3)}}\,,
\end{equation}
with
\begin{align}
Q_{\epsilon_{\text{lin}}} & =\oint d^{d-2}x\,\epsilon_{\text{lin}}\left(\pi_{(2)}^{r}-(d-3)\sqrt{\xbar g}\,\Psi^{(d-3)}\right)\,,\\ 
Q_{\xbar\epsilon} &=\oint d^{d-2}x\,\xbar\epsilon\left(\xbar\pi^{r}-(d-4)\sqrt{\xbar g}\,\Psi^{(d-4)}\right)\,,\label{eq:Qe}\\
Q_{\epsilon^{(d-4)}} & =-(d-4)\oint d^{d-2}x\sqrt{\xbar g}\,\epsilon^{(d-4)}\xbar\Psi\,, \quad
Q_{\epsilon^{(d-3)}}  =-(d-3)\oint d^{d-2}x\sqrt{\xbar g}\,\epsilon^{(d-3)}\Psi_{\text{lin}}\,,
\end{align}
and 
\begin{align}
Q_{\mu_{\text{lin}}} & =-\oint d^{d-2}x\sqrt{\xbar g}\,\mu_{\text{lin}}A_{r}^{(2)}\,,
&Q_{\xbar\mu} & =-\oint d^{d-2}x\sqrt{\xbar g}\,\xbar\mu\,\xbar A_{r}\,,\\
Q_{\mu^{(d-4)}} & =(d-4)\oint d^{d-2}x\sqrt{\xbar g}\,\mu^{(d-4)}\,\xbar \Phi\,,
&Q_{\mu^{(d-3)}} & =(d-3)\oint d^{d-2}x\sqrt{\xbar g}\,\mu^{(d-3)}\Phi_{\text{lin}}\,.
\end{align}

Notably, as in the seminal work by Brown and Henneaux in the case of AdS$_3$ gravity \cite{Brown:1986nw}, the Poisson brackets of the canonical generators acquire non trivial central charges
\begin{align}
\{G_{\epsilon_{\text{lin}}},G_{\mu^{(d-3)}}\} & =-\{G_{\mu^{(d-3)}},G_{\epsilon_{\text{lin}}}\}=-(d-3)\oint d^{d-2}x\sqrt{\xbar g}\,\epsilon_{\text{lin}}\,\mu^{(d-3)}\,,\\
\{G_{\xbar\epsilon},G_{\mu^{(d-4)}}\} & =-\{G_{\mu^{(d-4)}},G_{\xbar\epsilon}\}=-(d-4)\oint d^{d-2}x\sqrt{\xbar g}\,\xbar\epsilon\,\mu^{(d-4)}\,,\\
\{G_{\epsilon^{(d-4)}},G_{\xbar\mu}\} & =-\{G_{\xbar\mu},G_{\epsilon^{(d-4)}}\}=-(d-4)\oint d^{d-2}x\sqrt{\xbar g}\,\epsilon^{(d-4)}\,\xbar\mu\,,\\
\{G_{\epsilon^{(d-3)}},G_{\mu_{\text{lin}}}\} & =-\{G_{\mu_{\text{lin}}},G_{\epsilon^{(d-3)}}\}=-(d-3)\oint d^{d-2}x\sqrt{\xbar g}\,\epsilon^{(d-3)}\,\mu_{\text{lin}}\,.
\end{align}
Similar central extensions can also be found in the context of supergravity \cite{Fuentealba:2021xhn} and gravity in five dimensions \cite{Fuentealba:2022yqt} (in addition to the aforementioned four-dimensional cases of gravity \cite{Fuentealba:2022xsz} and electromagnetism \cite{Fuentealba:2023rvf}).

\vskip 2mm

Some comments are in order:
\begin{itemize}
\item  Preservation of the parity conditions on the fields implies that $\epsilon_{\text{lin}}$ is odd and $\mu_{\text{lin}}$ is even. Thus, one expects that the charge $Q_{\epsilon_{\text{lin}}}$, related to the electric dipole moment (for the $\ell=1$ component in the spherical harmonic expansion), together with its pair $Q_{\mu_{\text{lin}}}$ should combine to form an arbitrary angle-depedent $u(1)$ symmetry at null infinity. The latter combination of the charges should be connected to the Ward identities associated to the higher-dimensional subleading soft photon theorems found in \cite{He:2019pll}. As in the $d=4$ case (for the standard $u(1)$ symmetry), matching conditions at null infinity for the $\mathcal O(r)$ gauge symmetries are expected to be a consequence of the asymptotic behaviour at spatial infinity.

\item  From the expressions for the charges, one can also see that only the even part of $\epsilon^{(d-3)}$ and the odd part of $\mu^{(d-3)}$ generate improper gauge transformations. Thus, this pair should also combine in an arbitrary subleading $u(1)$ symmetry at null infinity. 

\item In spite of $\xbar \pi^r$ is a parity even function (as requested by parity conditions), both parities associated to the gauge parameter $\xbar \epsilon$ generate improper gauge transformations (unlike the four-dimensional case where only the even part of $\xbar \epsilon$ generates a non zero charge), and this is because of the parity odd field $\Psi^{(d-4)}$, which appears in the expression for the charge $Q_{\xbar \epsilon}$ in \eqref{eq:Qe}. A similar observation can be made for the parameter $\xbar \mu$, where both parities (of $\xbar \mu$) generate improper gauge transformations in all $d>4$, and in the $d=4$ case, it is only the parity odd part which span improper gauge symmetries (since $\xbar A_r$ becomes purely odd in $d=4$, as can be seen in \eqref{eq:Parity-cond-A-core}). Indeed, if we put $d=4$, the set of charges ($G_{\epsilon^{(d-4)}}$, $G_{\mu^{(d-4)}}$) vanishes, recovering the results found in \cite{Fuentealba:2023rvf}, but in absence of the logarithmic growth.

\item As aforementioned, for $d>4$ the parameters ($\xbar \epsilon$, $\xbar \mu$, $\epsilon^{(d-4)}$ and $\mu^{(d-4)}$) are not subject to any additional condition, generating all improper gauge symmetries. This explains why instead of having eight towers of infinite-dimensional abelian symmetries, the system is actually invariant under a six-fold set of angle-dependent $u(1)$ transformations.

\item We stress again that in the case without the linear growth in the gauge potential, the presence of the subleading field $\Psi^{(d-4)}$ of the radial expansion of $A_0$ in the symplectic \eqref{eq:Omega0} form renders improper the subleading gauge transformations generated by $\epsilon^{(d-4)}$ and $\mu^{(d-4)}$. These new improper symmetries enable us to obtain central charges between the standard $u(1)$ canonical generators ($G_{\xbar \epsilon}$, $ G_{\xbar \mu}$) and the subleading ones ($G_{\epsilon^{(d-4)}}$, $ G_{\mu^{(d-4)}}$). A similar effect occurs when one relaxes the asymptotic conditions with the linear growth. The term along the subleading field $\Psi^{(d-3)}$ (of the radial expansion of $A_0$) in the boundary integral of the symplectic form \eqref{eq:Omega0} makes improper the subleading gauge transformations generated by $\epsilon^{(d-3)}$ and $\mu^{(d-3)}$, finding hence central charges betwee the  canonical generators ($G_{\epsilon_{\text{lin}}}$, $ G_{\mu_{\text{lin}}}$) and the subleading generators ($G_{\epsilon^{(d-3)}}$, $G_{\mu^{(d-3)}}$). 
\end{itemize}

\section{Canonical realization of the Poincar\'e symmetry}\label{sec:Section4}

In order to obtain a canonical action of the Poincar\'e symmetry, the symplectic form must be invariant, i.e., $\mathcal L_{X_{\xi,\xi^i}} \Omega=0$. As noted in the case of the $u(1)$ gauge transformations treated in the latter section, this means that $X_{\xi,\xi^i}$ is a Hamiltonian vector field generating canonical phase space transformations, which implies 
\begin{equation}
\iota_{X_{\xi,\xi^i}} \Omega=-d_V P_{\xi,\xi^i}\,,
\end{equation} 
where $P_{\xi,\xi^i}$ is the canonical generator. The Poincar\'e Killing vector in polar coordinates reads
\begin{equation}
\xi=br+T\,,\quad \xi^r=W\,,\quad \xi^A=Y^A+\frac{1}{r}\xbar D^AW\,,
\end{equation}
where the boost parameter $b$ and the spatial translations $W$ are both in the kernel of the differential operator $(\xbar D_A\xbar D_B+\xbar g_{AB})$, while time translations satisfy that $\partial_A T=0$. 

As we explicitly show in Appendix \ref{app:AppendixA}, the purely bulk symplectic form fails to be Poincar\'e invariant. This can be cured by adding a very precise boundary term to the symplectic form (which explains the boundary term $\mathcal S$ in \eqref{eq:S}) and applying additional corrective gauge transformations (by finding $\epsilon_{(\xi,\xi^i)}$ and $\mu_{(\xi,\xi^i)}$ in \eqref{eq:A_i} and \eqref{eq:A_0}, respectively). Parity conditions play a key role by taking care of divergences in $\mathcal L_{X_{\xi,\xi^i}} \Omega$ (see Appendix \ref{app:AppendixA} for details). 

We find that the corrective parameters are asymptotically given by
\begin{align}
\epsilon_{(\xi,\xi^i)} & =\frac{1}{r^{d-4}}\epsilon^{(d-4)}_{(T,W)}+\frac{1}{r^{d-3}}\epsilon^{(d-3)}_{(T,W)}+\mathcal{O}\left(r^{-d+2}\right)\,,\\
\mu_{(\xi,\xi^i)} & =\frac{1}{r^{d-3}}\mu^{(d-4)}_{(b,T,W)}+\frac{1}{r^{d-2}}\mu^{(d-3)}_{(b,T,W)}+\mathcal{O}\left(r^{-d+1}\right)\,,
\end{align}
where
\begin{align}
\epsilon^{(d-4)}_{(T,W)}&=-T\Psi^{(d-5)}-\partial_A W \xbar D^A \Phi^{(d-5)}+(d-5)W\Phi^{(d-5)}\,,\\
\epsilon^{(d-3)}_{(T,W)}&=\frac{1}{d-3}\left(-T\Psi^{(d-4)}-\partial_A W \xbar A^A+W\xbar A_r\right)\,,\\
\mu^{(d-4)}_{(b,T,W)}&=-\left[2b\xbar A_r+T\left(\xbar \triangle \Phi^{(d-5)}-2(d-5)\Phi^{(d-5)}\right)+\partial_A W\xbar D^A\Psi^{(d-5)}-(d-4)W\Psi^{(d-5)}\right]\,,\\
\mu^{(d-3)}_{(b,T,W)}&=\frac{1}{d-3}\Big[-2(d-2)bA^{(2)}_r-T\xbar D_A \xbar A^A-(2d-5)T\xbar A_r\nonumber\\
&\qquad \qquad \,\,\, -\frac{1}{\sqrt{\xbar g}}\partial_A W \Big(\xbar \pi^A+\sqrt{\xbar g}\xbar D^A \Psi^{(d-4)}\Big)+\frac{W}{\sqrt{\xbar g}}\Big(\xbar \pi^r-(d-5)\sqrt{\xbar g}\Psi^{(d-4)}\Big)\Big]\,.
\end{align}
Note that these field-dependent corrective gauge transformations (required by integrability of the canonical generators) could lead to modifications of the transformation laws of the fields in equations \eqref{eq:A_i}, \eqref{eq:pi_i}, \eqref{eq:A_0} and \eqref{eq:pi_0}, by constraint terms. However, due to their fast decay, these terms do not change our results.

Using the transformation laws of the fields listed in appendix \ref{app:AppendixB}\,, we obtain that the canonical generator of Poincar\'e symmetry is given by
\begin{equation}
P_{\xi,\xi^i}=\int d^{d-1}x \left(\xi \mathcal H+\xi^i\mathcal H_i+\epsilon_{(\xi,\xi^i)}\mathcal G+ \mu_{(\xi,\xi^i)}\pi^0\right)+\mathcal Q_{\xi,\xi^i}\,,
\end{equation}
where $\mathcal H$ is given in \eqref{eq:Ham-density} and
\begin{align}
\mathcal{H}_{i} & =F_{ij}\pi^{j}-\partial_{j}\pi^{j}A_{i}+\pi^0\partial_{i}A_0\,,\\
\mathcal Q_{\xi,\xi^{i}} & =\oint d^{d-2}x\Big\{
 T\Big(\xbar\Pi^{r}\Psi_{\text{lin}}+\sqrt{\xbar g}\,\partial_{A}\xbar A_{r}\xbar D^{A}\Phi_{\text{lin}}-(d-2)\sqrt{\xbar g}\,\xbar A_{r}\Phi_{\text{lin}}\Big)\\
 & \quad+W\Big[\xbar D_A(\xbar D^A \xbar \Pi^r\Phi_{\text{lin}})+(d-1)\xbar \Pi^r\Phi_{\text{lin}}+\sqrt{\xbar g}\xbar D_A(\xbar A_r\xbar D^A\Psi_{\text{lin}})\Big]\\
&\quad+ Y^{A}\left(\xbar\Pi^{r}\xbar D_{A}\xbar \Phi+\Pi_{(2)}^{r}\partial_{A}\Phi_{\text{lin}}+\sqrt{\xbar g}\,\xbar\Psi\partial_{A}\xbar A_{r}+\sqrt{\xbar g}\,\Psi_{\text{lin}}\partial_{A}A_{r}^{(2)}\right)\\  
&\quad+b\Big(\xbar\Pi^{r}\xbar\Psi+\Pi_{(2)}^{r}\Psi_{\text{lin}}+\sqrt{\xbar g}\partial_{A}\xbar A_{r}\xbar D^{A}\xbar \Phi+\sqrt{\xbar g}\,\partial_{A}A_{r}^{(2)}\xbar D^{A}\Phi_{\text{lin}}-(d-1)\sqrt{\xbar g}\,A_{r}^{(2)}\Phi_{\text{lin}}\Big)\Big\}\,,
\end{align}
with
\begin{align}
\xbar\Pi^{r} & =\xbar\pi^{r}-(d-4)\sqrt{\xbar g}\,\Psi^{(d-4)}\,,\\
\Pi_{(2)}^{r} & =\pi_{(2)}^{r}-(d-3)\sqrt{\xbar g}\,\Psi^{(d-3)}\,.
\end{align}
By making use of the transformation laws of the fields listed in Appendix \ref{app:AppendixB}, we find that the Poisson brackets of the canonical generators are given by
\begin{align}
\big\{ P_{\xi_{1},\xi_{1}^{i}},P_{\xi_{2},\xi_{2}^{i}}\big\} & =P_{\hat{\xi},\hat{\xi}^{i}}\,,\\
\big\{ G_{\mu,\epsilon},P_{\xi,\xi^{i}}\big\} & =G_{\hat{\mu},\hat{\epsilon}}\,,\\
\big\{ G_{\mu_{1},\epsilon_{1}},G_{\mu_{2},\epsilon_{2}}\big\} & =C_{\{\mu_{1},\epsilon_{1};\mu_{2},\epsilon_{2}\}}\,,\label{eq:GG}
\end{align}
where the Poincar\'e Killing vectors transform as
\begin{align}
\hat{\xi} & =\xi_{1}^{i}\partial_{i}\xi_{2}-\xi_{2}^{i}\partial_{i}\xi_{1}\,,\\
\hat{\xi}^{i} & =\xi_{1}^{j}\partial_{j}\xi_{2}^{i}-\xi_{2}^{j}\partial_{j}\xi_{1}^{i}+g^{ij}\left(\xi_{1}\partial_{j}\xi_{2}-\xi_{2}\partial_{j}\xi_{1}\right)\,,
\end{align}
and the transformation laws of the improper gauge parameters read
\begin{align}
\hat{\epsilon}_{\text{lin}} & =-Y^{A}\partial_{A}\epsilon_{\text{lin}}-b\mu_{\text{lin}}\,,\\
\hat{\mu}_{\text{lin}} & =-Y^{A}\partial_{A}\mu_{\text{lin}}-(d-1)b\epsilon_{\text{lin}}-\xbar D_{A}\left(b\xbar D^{A}\epsilon_{\text{lin}}\right)\,,\\
\hat{\xbar\epsilon} & =-Y^{A}\partial_{A}\xbar\epsilon-b\xbar\mu-T\mu_{\text{lin}}-W\epsilon_{\text{lin}}-\partial_{A}W\xbar D^{A}\epsilon_{\text{lin}}\,,\\
\hat{\xbar\mu} & =-Y^{A}\partial_{A}\xbar\mu-\xbar D_{A}\left(b\xbar D^{A}\xbar\epsilon\right)-T\left(\xbar\triangle\,\epsilon_{\text{lin}}+(d-2)\epsilon_{\text{lin}}\right)-\partial_{A}W\xbar D^{A}\mu_{\text{lin}}\,,\\
\hat{\epsilon}^{(d-4)} & =-Y^{A}\partial_{A}\epsilon^{(d-4)}-b\mu^{(d-4)}\,,\\
\hat{\mu}^{(d-4)} & =-Y^{A}\partial_{A}\mu^{(d-4)}-\xbar D_{A}\left(b\xbar D^{A}\epsilon^{(d-4)}\right)\,,\\
\hat{\epsilon}^{(d-3)} & =-Y^{A}\partial_{A}\epsilon^{(d-3)}-b \mu^{(d-3)}\nonumber\\
&\quad-\frac{(d-4)}{(d-3)}\left[T\mu^{(d-4)}+\partial_{A}W \xbar D^A\epsilon^{(d-4)}-(d-2)W\epsilon^{(d-4)}\right]\,,\\
\hat{\mu}^{(d-3)} & =-Y^{A}\partial_{A}\mu^{(d-3)}-(d-1)b\epsilon^{(d-3)}-\xbar D_{A}\left(b\xbar D^{A}\epsilon^{(d-3)}\right)\nonumber\\
&\quad-\frac{(d-4)}{(d-3)}\left[T\left(\xbar\triangle\,\epsilon^{(d-4)}+(d-2)\epsilon^{(d-4)}\right)+\partial_{A}W\xbar D^{A}\mu^{(d-4)}-(d-1)W\mu^{(d-4)}\right]\,.
\end{align}
The central charges $C_{\{\mu_{1},\epsilon_{1};\mu_{2},\epsilon_{2}\}}$, that appear in the bracket between the gauge canonical generators in \eqref{eq:GG}, can be found in Section \ref{sec:Section3}. 

One can check that all the Jacobi identities are identically satisfied.

The charges associated to $(Q_{\epsilon_{\text{lin}}},Q_{\mu_{\text{lin}}})$ and $(Q_{\epsilon^{(d-4)}},Q_{\mu^{(d-4)}})$ (which play the role of the logarithmic $u(1)$ symmetries in $d=4$) are not conserved in time (under the assumption of non explicit time dependence). This can be readily checked from the integrand of these charges (given in Section \ref{sec:Section3}) and using the equations of motion found in Section \ref{sec:Section2}, specifically in \eqref{eq:dipole-moment}, \eqref{eq:E2}, \eqref{eq:A02} and \eqref{eq:Phi-dot}, respectively. This is indeed consistent with the brackets of these canonical generators with the Hamiltonian $H$ (as can be directly read from the transformation law of the parameters). Similar phenomenon occurs in the case of $d=4$, where linear and logarithmic $u(1)$ charges are not conserved in time (with no explicit time dependence on the parameters) \cite{Fuentealba:2023rvf}.

As we shall see in the next section, this specific algebraic structure allows to redefine the Poincar\'e canonical generators, so that all of the $u(1)$ charges transform in the trivial representation of the Lorentz group. The direct sum form of the algebra has as direct consequence the conservation in time of all $u(1)$ charges.

\section{Gauge-invariant Poincar\'e generators }\label{sec:Section5}

The general argument in \cite{Fuentealba:2022xsz} gives a mechanism to disentangle all pure supertranslations from the Poincar\'e algebra through a very precise nonlinear automorphism\footnote{By nonlinear automorphism we refer to a nonlinear map of the generators that preserves a given Lie algebra. In the case of infinite-dimensional Lie algebras this sort of redefinitions has shown to play a central role in the mapping between different two-dimensional models \cite{Rodriguez:2021tcz,Tempo:2022ndz,Bagchi:2022nvj}. See also Section 5.1 in \cite{Fuentealba:2022yqt} for a recent discussion on nonlinear redefinitions (of Lie algebra generators) and Poisson manifolds.} of the Poincar\'e canonical generators. This procedure works due to the existence of invertible central charges in the abelian algebra of all supertranslations, for which logarithmic supertranslations were crucial (central charges appear in the bracket between standard and logarithmic supertranslations). 

The mechanism was then applied to the case of electromagnetism in \cite{Fuentealba:2023rvf}, which it was possible to carry out due to the relaxation of the asymptotic conditions by logarithmic terms in the gauge potential. Again, the presence of invertible central charges, allowed us to perform an appropriate nonlinear Poincar\'e  automorphism that decouple all the $u(1)$ symmetries from the Poincar\'e algebra. Finding in this way a definition for the angular momentum free of $u(1)$ ambiguities (in line with the Coleman-Mandula theorem \cite{Coleman:1967ad}).

Given the form of the asymptotic symmetry algebra found in this paper (with invertible central charges), we can find the precise nonlinear automorphism of the Poincar\'e algebra. This is implemented through a suitable improper gauge transformation with the parameters $\epsilon^{\text{extra}}_{(\xi,\xi^i)}$ and $\mu^{\text{extra}}_{(\xi,\xi^i)}$, such that the new Poincar\'e generator 
\begin{equation}
\tilde{P}_{\xi,\xi^i}=\int d^{d-1}x \left(\xi \mathcal H+\xi^i\mathcal H_i+\epsilon_{(\xi,\xi^i)}\mathcal G \mu_{(\xi,\xi^i)} \pi^0\right)+\mathcal Q_{\xi,\xi^i}+ G_{\epsilon^{\text{extra}}_{(\xi,\xi^i)},\mu^{\text{extra}}_{(\xi,\xi^i)}}\,,
\end{equation}
satisfies the following Poisson brackets
\begin{eqnarray}
\big\{\tilde{P}_{\xi_{1},\xi_{1}^{i}},\tilde{P}_{\xi_{2},\xi_{2}^{i}}\big\}&=&\tilde{P}_{\hat{\xi},\hat{\xi}^{i}}\,,\\
\big\{ G_{\mu,\epsilon},\tilde{P}_{\xi,\xi^{i}}\big\}&=&0\,.
\end{eqnarray}
We find that the extra surface integral in $G_{\epsilon^{\text{extra}}_{(\xi,\xi^i)},\mu^{\text{extra}}_{(\xi,\xi^i)}}$ reads
\begin{align}
\mathcal Q^{\text{extra}}_{\xi,\xi^i} & =-\oint d^{d-2}x\Big\{T\Big(\xbar\Pi^{r}\Psi_{\text{lin}}+\sqrt{\xbar g}\,\partial_{A}\xbar A_{r}\xbar D^{A}\Phi_{\text{lin}}-(d-2)\sqrt{\xbar g}\,\xbar A_{r}\Phi_{\text{lin}}\Big)\\
 & \quad+W\Big[\xbar D_A(\xbar D^A \xbar \Pi^r\Phi_{\text{lin}})+(d-1)\xbar \Pi^r\Phi_{\text{lin}}+\sqrt{\xbar g}\xbar D_A(\xbar A_r\xbar D^A\Psi_{\text{lin}})\Big]\\
&\quad+ Y^{A}\left(\xbar\Pi^{r}\xbar D_{A}\xbar \Phi+\Pi_{(2)}^{r}\partial_{A}\Phi_{\text{lin}}+\sqrt{\xbar g}\,\xbar\Psi\partial_{A}\xbar A_{r}+\sqrt{\xbar g}\,\Psi_{\text{lin}}\partial_{A}A_{r}^{(2)}\right)\\  
&\quad+b\Big(\xbar\Pi^{r}\xbar\Psi+\Pi_{(2)}^{r}\Psi_{\text{lin}}+\sqrt{\xbar g}\partial_{A}\xbar A_{r}\xbar D^{A}\xbar \Phi+\sqrt{\xbar g}\,\partial_{A}A_{r}^{(2)}\xbar D^{A}\Phi_{\text{lin}}-(d-1)\sqrt{\xbar g}\,A_{r}^{(2)}\Phi_{\text{lin}}\Big)\Big\}\,.
\end{align}
The asymptotic behaviour of the field-dependent parameters $\epsilon^{\text{extra}}_{(\xi,\xi^i)}$ and $\mu^{\text{extra}}_{(\xi,\xi^i)}$ generating the above charge, is determined by the following parameters
\begin{align}
\epsilon_{\text{lin}} & =-\mathcal L_Y \Phi_{\text{lin}}-b\Psi_{\text{lin}}\,,\\
\mu_{\text{lin}} & =-\mathcal L_Y \Psi_{\text{lin}}-\left[\xbar D_A (b\xbar D^A \Phi_{\text{lin}})+(d-1)b \Phi_{\text{lin}}\right]\,,\\
\xbar\epsilon & =-\mathcal L_Y \xbar \Phi-b\xbar \Psi-T\Psi_{\text{lin}}-\partial_AW\xbar D^A\Phi_{\text{lin}}-W\Phi_{\text{lin}}\,,\\
\xbar\mu  & =-\mathcal L_Y \xbar \Psi-\xbar D_A (b\xbar D^A \xbar \Phi)-T\left(\xbar \triangle\,\Phi_{\text{lin}}+(d-2)\Phi_{\text{lin}}\right)-\partial_A W\xbar D^A \Psi_{\text{lin}}\,,\\
\epsilon^{(d-4)} & =\frac{1}{d-4}\left(\mathcal L_Y \xbar A_r+\frac{b\xbar \Pi^r}{\sqrt{\xbar g}}\right)\,,\\
\mu^{(d-4)} & =\frac{1}{d-4}\left[\frac{1}{\sqrt{\xbar g}}\mathcal L_Y \xbar \Pi^r+\xbar D_A(b\xbar D^A \xbar A_r)\right]\,,\\
\epsilon^{(d-3)} & =\frac{1}{d-3}\left[\mathcal L_Y A^{(2)}_r+\frac{1}{\sqrt{\xbar g}}b\Pi^r_{(2)}+\frac{1}{\sqrt{\xbar g}}T\xbar \Pi^r+\xbar D_A(\xbar D^A W\xbar A_r)\right]\,,\\
\mu^{(d-3)} & =\frac{1}{d-3}\left[\frac{1}{\sqrt{\xbar g}}\mathcal L_Y \Pi^r_{(2)}+\xbar D_A(b\xbar D^A A^{(2)}_r)+(d-1)b A^{(2)}_r+T\left(\xbar \triangle\,\xbar A_r+(d-2)\xbar A_r\right)\right]\nonumber\\
&\quad+\frac{1}{(d-3)\sqrt{\xbar g}}\left(\partial_A \xbar D^A W \xbar \Pi^r-(d-1)W\xbar \Pi^r\right) \,.
\end{align}
Thus, the gauge-invariant Poincar\'e generator reads
\begin{equation}
\tilde{P}_{\xi,\xi^i}=\int d^{d-1}x \left[\xi \mathcal H+\xi^i\mathcal H_i+(\epsilon_{(\xi,\xi^i)}+\epsilon^{\text{extra}}_{(\xi,\xi^i)})\mathcal G+ (\mu_{(\xi,\xi^i)}+\mu^{\text{extra}}_{(\xi,\xi^i)})\pi^0\right]\,.
\end{equation}
We have thus shown that the decoupling of the $u(1)$ charges from the Poincar\'e algebra is a general property of electromagnetism in \textit{all} spacetime dimensions $d\geq 4$.

\section{Concluding remarks}\label{sec:Section6}

In this paper, we have extended the results recently obtained in \cite{Fuentealba:2023rvf} to all higher spacetime dimensions. In particular, we have shown how to consistently relax the asymptotic conditions in electromagnetism, by accommodating a linear growth of the gauge potential at spatial infinity. The linear relaxation was done by adopting a different set of boundary conditions with respect to the ones in \cite{Henneaux:2019yqq}, which leads to a slower decay of the magnetic field. Thus, solutions that possess magnetic sources with monopole decay at infinity are included. Consistency of Poincar\'e invariance with finiteness of the symplectic structure imposes a set of strict parity conditions on the leading order fields. This is in contrast with the analysis performed in \cite{Henneaux:2019yqq} where no parity conditions were needed. Notably, the boundary conditions introduced in this work render improper a set of subleading gauge symmetries, which mimic the subleading supertranslations found in the context of gravity in higher dimensions \cite{Fuentealba:2021yvo,Fuentealba:2022yqt}. The asymptotic symmetries are given by a six-fold family of angle-dependent $u(1)$ transformations. We have also found that the canonical generators associated to these gauge symmetries obey an abelian algebra with non zero central charges. Finally, we have made use of the mechanism proposed in \cite{Fuentealba:2022xsz} to decouple all the $u(1)$ gauge transformations from the Poincar\'e algebra. This is realized through a nonlinear automorphism of the Poincar\'e algebra, which makes the canonical generators of the improper gauge symmetries Poincar\'e scalars.

One could wonder if extensions with higher powers of $r$ or even containing a logarithmic branch can be accommodated in the asymptotic condition of the gauge field. Nonetheless, preliminary results suggest that the former growth would lead to divergences in the symplectic structure which are difficult to cure by using parity conditions. The logarithmic growth on the other hand does not seem to contribute to the surface integral of the canonical generator, being associated then to a proper gauge symmetry (in $d>4$). Thus, a connection to a would-be higher-dimensional generalization of the results obtained in \cite{Peraza:2023ivy} seems to be non trivial. Indeed, it would be of interest to analyse to what extent this sort of symmetries emerges either in the light-cone formulation for electromagnetism in higher dimensions, along the lines of \cite{Majumdar:2022fut}, or in the light-front (where the flat space is foliated by retarded/advanced time) \cite{Gonzalez:2023yrz}. This could shed some light on how to understand (or even circumvent) the obstructions, that we have faced so far, in order to find further consistent relaxations (with higher powers of $r$) respecting both finiteness and exact invariance of the symplectic form. These requirements are essential to guarantee the direct application of standard Hamiltonian methods to obtain well-defined canonical generators, whose algebra does satisfy all the necessary Jacobi identities by construction.

It must be emphasized that the linear behaviour of the gauge potential requires, in addition to the parity conditions on the leading order fields, to impose twisted parity conditions on the ``core'' fields (which are associated to the Weinberg's leading soft theorems). Thus, it would be also interesting to connect these results to null infinity, where the necessary matching conditions in  \cite{He:2019pll,He:2019jjk} are expected to be consequence of the asymptotic behaviour at spatial infinity. This particular problem for all dimensions is harder (due to the more intricate definition of null infinity in odd dimensions). Work along these lines is currently in progress \cite{Fuentealba:matchHD}.

\section*{Acknowledgments}
I would like to express my gratitude to Marc Henneaux and Ricardo Troncoso for their careful reading of this manuscript and valuable comments. Special thanks to Marc Henneaux for his encouragement to write this paper. I would also like to thank Jos\'e Figueroa, Hern\'an Gonz\'alez, Jakob Salzer and  C\'edric Troessaert for discussions. This work was partially supported by a Marina Solvay Fellowship,  FNRS-Belgium (conventions FRFC PDRT.1025.14 and IISN 4.4503.15), as well as by funds from the Solvay Family.

\appendix

\section{Solving the non-integrability of the Poincar\'e generators}
\label{app:AppendixA}

In this appendix, we shall derive, in a constructive way, the boundary deformation of the symplectic form together with the necessary corrective gauge transformations (along the Poincar\'e parameters) in order to obtain a canonical action of the Poincar\'e symmetry ($\mathcal L_{\xi,\xi^i}\Omega=0$) for the relaxed set of asymptotic conditions presented in this work. 

Let us first consider the pure bulk symplectic form for the usual Hamiltonian theory,
\begin{equation}
\Omega^{\text{bulk}}=\int d^{d-1}x d_{V}\pi^{i}d_{V}A_{i}\,.
\end{equation}
The need of considering the extended Hamiltonian formulation will be explicit in the course of the calculation. This is due to the fact that $A_0$ carries boundary degrees of freedom, which are the key ingredients for having boundary conditions that are invariant under both Poincar\'e and angle-dependent $u(1)$ transformations. 

In what follows, it is computed the change in the symplectic form by acting with a normal Poincar\'e diffeormorphism $\xi=br+T$. This suggests the boundary term that must be added to the bulk symplectic form (associated to $\mathcal S$) and the additional boundary degrees of freedom (ultimately related to $A_0$). As anticipated, additional corrective gauge transformations must be performed in order to obtain that $\mathcal L_{X_{(\xi,\xi^i)}}\Omega=0$, where $\xi^i$ is the spatial Poincar\'e deformation $\xi^i=(\xi^r,\xi^A)=(W,Y^A+\xbar D^A W/r)$.

\subsection*{Change of $\Omega^{\text{bulk}}$ along normal Poincar\'e transformations}
By using the transformation laws
\begin{align}
\delta_{\xi}A_{i} & =\frac{\xi\pi_{i}}{\sqrt{g}}\,,\\
\delta_{\xi}\pi^{i} & =\sqrt{g}\nabla_{m}\big(F^{mi}\xi\big)\,,
\end{align}
we obtain that the variation of the bulk symplectic form under $\xi$ becomes
\begin{align}
\mathcal L_{\xi}\Omega^{\text{bulk}}&=\int d^{d-2}S_i\xi d_V F^{ij} d_V A_j\\
&=\oint d^{d-2}x\sqrt{\xbar g}\,r^{d-4}\left(br+T\right)\xbar g^{AB}d_{V}F_{rA}d_{V}A_{B}\,.
\end{align}
The asymptotic conditions for the gauge potential imply that
\begin{equation}
F_{rA}=-\frac{1}{r^{d-3}}\left(\partial_{A}\xbar A_{r}+(d-4)\xbar A_{A}\right)-\frac{1}{r^{d-2}}\left(\partial_{A}A_{r}^{(2)}+(d-3)A_{A}^{(2)}\right)+\mathcal O\left(r^{-d+1}\right)\,.
\end{equation}
The variation of $\Omega^{\text{bulk}}$ then becomes
\begin{align}
\mathcal{L}_{\xi}\Omega^{\text{bulk}}  =&-r\oint d^{d-2}x\sqrt{\xbar g}\,b \,d_{V}\left(\partial_{A}\xbar A_{r}+(d-4)\xbar A_{A}\right)\xbar D^{A}d_{V}\Phi_{\text{lin}}\nonumber\\
&+\oint d^{d-2}x\sqrt{\xbar g}\,d_{V}\xbar A_{r}\left[\partial_{A}\left(bd_{V}\xbar D^{A}\Phi\right)+T\xbar\triangle\,d_{V}\Phi_{\text{lin}}\right]\nonumber\\
&+(d-4)\oint d^{d-2}x\sqrt{\xbar g}\,\partial_{A}\left(bd_{V}\xbar A^{A}\right)d_{V}\Phi\nonumber \\
& \quad+\oint d^{d-2}x\sqrt{\xbar g}\,d_{V}A_{r}^{(2)}\partial_{A}\left(b\xbar D^{A}d_{V}\Phi_{\text{lin}}\right)\nonumber\\
&+(d-3)\oint d^{d-2}x\sqrt{\xbar g}\left[\partial_{A}\left(bd_{V}A^{(2)A}\right)+\left(\frac{d-4}{d-3}\right)T\xbar D_{A}d_{V}\xbar A^{A}\right]d_{V}\Phi_{\text{lin}}\,.\label{eq:old-variation}
\end{align}
By making use of the parity conditions on the fields declared in section \ref{sec:Section2}, we can eliminate the linear divergence that appears in \eqref{eq:old-variation}  (the integrand becomes an odd function under the antipodal map). 

The remaining finite terms in \eqref{eq:old-variation} are not zero and they cannot be eliminated by corrective gauge transformations. We shall solve this problem by following the approach developed in \cite{Henneaux:2018gfi,Henneaux:2019yqq}, namely, by introducing new boundary degrees of freedom and deforming the symplectic form by boundary terms. 

We adopt in this point a different set of boundary conditions as the ones considered in \cite{Henneaux:2019yqq}. To show how this works, let us switch off the linear fields for the moment. In \cite{Henneaux:2019yqq} the problem was solved by introducing a new boundary field $\xbar \Psi$ (related to $A_0$) and imposing that $\xbar A_A=\xbar D_A \Theta$ (equivalent to a faster fall-off of the magnetic field), which makes the formalism to work perfectly, making the Lorentz boosts canonical transformations (see \cite{Henneaux:2019yqq}). However, we propose here another solution, by leaving $\xbar A_A$  arbitrary, but at the price to introduce a new  boundary field $\Psi^{(d-4)}$. The latter is nothing but a subleading term in the radial expansion of $A_0$ (see below). Now, let us switch on the linear behaviour again. To solve integrability problems, we introduce the new boundary fields  $\Psi_{\text{lin}}$ and $\Psi^{(d-3)}$, such that the asymptotic behaviour  of $A_0$ is given by
\begin{equation}\label{eq:A0}
A_0  =\Psi_{\text{lin}}+\frac{1}{r}\xbar\Psi+\dots+\frac{1}{r^{d-3}}\Psi^{(d-4)}+\frac{1}{r^{d-2}}\Psi^{(d-3)}+\mathcal{O}\left(r^{-d+1}\right)\,.
\end{equation}

\subsection*{Symplectic form for the extended phase space}

We shall show now the invariance of the system under the Poincar\'e symmetry. For this we extend the phase space by the conjugate pair  $(A_0,\pi^0)$, and consider the following symplectic form
\begin{align}
\Omega & =\int d^{d-1}x\big(d_{V}\pi^{i}d_{V}A_{i}+d_{V}\pi^0 d_{V}A_0\big)\nonumber\\
 & \quad-\oint d^{d-2}x\sqrt{\xbar g}\Big[d_{V}\xbar A_{r}d_{V}\xbar\Psi+(d-4)d_{V}\Psi^{(d-4)}d_{V}\xbar \Phi \nonumber\\
 &\qquad \qquad \qquad \qquad+d_{V}A_{r}^{(2)}d_{V}\Psi_{\text{lin}}+(d-3)d_{V}\Psi^{(d-3)}d_{V}\Phi_{\text{lin}}\Big]\,.
\end{align}
The form of the additional boundary term is suggested by the equation \eqref{eq:old-variation}. In order to show the invariance of the theory, we apply the transformation rules already proposed in \cite{Henneaux:2018gfi,Henneaux:2019yqq}, to wit,
\begin{align}
\delta_{\xi}A_{i} & =\frac{\xi\pi_{i}}{\sqrt{g}}+\partial_{i}\left(\xi A_0 \right)\,, \\
\delta_{\xi}\pi^{i} & =\sqrt{g}\nabla_{m}\big(F^{mi}\xi\big)+\xi\partial^{i}\pi^0\,,\\
\delta_{\xi}A_0 & =\nabla_{i}\left(\xi A^{i}\right)\,,\\
\delta_{\xi}\pi^0 & =\xi\partial_{i}\pi^{i}\,, \label{eq:Poinc4}
\end{align}
which imply the following transformation laws for the boundary fields
\begin{align}
\delta_{\xi}\Psi_{\text{lin}} & =\xbar D_{A}\left(b\xbar D^{A}\Phi_{\text{lin}}\right)+(d-1)b\Phi_{\text{lin}}\,,\\
\delta_{\xi}\xbar\Psi & =\xbar D_{A}\big(b\xbar D^{A}\xbar\Phi\big)+T\left(\xbar\triangle\,\Phi_{\text{lin}}+(d-2)\Phi_{\text{lin}}\right)\,,\\
\delta_{\xi}\Psi^{(d-4)} & =\xbar D_{A}\left(b\xbar A^{A}\right)+2b\xbar A_{r}+T\left(\xbar\triangle\,\Phi^{(d-5)}-2(d-5)\Phi^{(d-5)}\right)\,,\\
\delta_{\xi}\Psi^{(d-3)} & =\xbar D_{A}\left(bA^{(2)A}\right)+bA_{r}^{(2)}+T\left(\xbar D_{A}\xbar A^{A}+\xbar A_{r}\right)\,,
\end{align}
\begin{align}
\delta_{\xi}\Phi_{\text{lin}} & =b\Psi_{\text{lin}}\,,\\
\delta_{\xi}\xbar \Phi & =b\xbar\Psi+T\Psi_{\text{lin}}\,,\\
\delta_{\xi}\xbar A_{r} & =\frac{b}{\sqrt{\xbar g}}\xbar\pi_{r}-(d-4)b\Psi^{(d-4)}-(d-4)T\Psi^{(d-5)}\,,\\
\delta_{\xi}A_{r}^{(2)} & =\frac{b}{\sqrt{\xbar g}}\pi_{(2)}^{r}-(d-3)b\Psi^{(d-3)}+\frac{T}{\sqrt{\xbar g}}\xbar\pi^{r}-(d-3)T\Psi^{(d-4)}\,.
\end{align}

A direct calculation shows that
\begin{align}
\mathcal L_{\xi}\Omega =\oint d^{d-2}x\sqrt{\xbar g} \Big\{&(d-4)T d_V \Psi^{(d-5)} d_V \xbar \Psi +T d_V \Psi^{(d-4)}d_V\Psi_{\text{lin}} \nonumber\\
&-(d-4)\Big[2b d_V\xbar A_r+T\Big(\xbar \triangle\, d_V\Phi^{(d-5)}-2(d-5)d_V\Phi^{(d-5)}\Big)\Big]d_V\bar \Phi \nonumber \\
&-\Big[2(d-2)bd_V A^{(2)}_r+T\xbar D_Ad_V\xbar A^A+(2d-4)Td_V \xbar A_r\Big]d_V \Phi_{\text{lin}}\Big\}\,.
\end{align}
These terms can now be cancelled out by additional corrective gauge transformation. In this case we must perform subleading gauge transformations whose parameters read
\begin{align}
\epsilon^{(d-4)}_{(T)}&=-T\Psi^{(d-5)}\,,\\
\epsilon^{(d-3)}_{(T)}&=\frac{1}{d-3}\left[-T\Psi^{(d-4)}\right]\,,\\
\mu^{(d-4)}_{(b,T)}&=-\left[2b\xbar A_r+T\left(\xbar \triangle \Phi^{(d-5)}-2(d-5)\Phi^{(d-5)}\right)\right]\,,\\
\mu^{(d-3)}_{(b,T)}&=\frac{1}{d-3}\Big[-2(d-2)bA^{(2)}_r-T\xbar D_A \xbar A^A-(2d-5)T\xbar A_r \Big]\,.
\end{align}
Thus, the change of the new symplectic form under $\xi$ is zero, i.e., $\mathcal L_{\xi}\Omega=0$.

On the other hand, the change of the new symplectic form $\Omega$ under spatial diffeomorphisms $\xi^i$ reveals that additional corrective subleading gauge transformations along spatial translations $W$ must be performed. Thus, we obtain that
\begin{align}
\mathcal L_{\xi^i}\Omega =\oint d^{d-2}x\sqrt{\xbar g} \Big\{&(d-4)\Big[\partial_A W\xbar D^A d_V\Phi^{(d-5)}-(d-5)W d_V\Phi^{(d-5)}\Big] d_V \xbar \Psi \nonumber\\
&+\Big(\partial_A Wd_V\xbar A^A-Wd_V\xbar A_r\Big)d_V\Psi_{\text{lin}} \nonumber\\
&-(d-4)\Big[\partial_A W\xbar D^Ad_V\Psi^{(d-5)}-(d-4)Wd_V\Psi^{(d-5)}\Big]d_V\bar \Phi \nonumber \\
&+\frac{1}{\sqrt{\xbar g}}\Big[Wd_V\Big(\xbar \Pi^r-\sqrt{\xbar g}\Psi^{(d-4)}\Big)-\partial_AW d_V\Big(\xbar\pi^A+\sqrt{\xbar g}\xbar D^A\Psi^{(d-4)}\Big) \Big]d_V \Phi_{\text{lin}}\Big\}\,.
\end{align}
which is identically cancelled out by gauge transformations with parameters
\begin{align}
\epsilon^{(d-4)}_{(W)}&=-\partial_A W \xbar D^A \Phi^{(d-5)}+(d-5)W\Phi^{(d-5)}\,,\\
\epsilon^{(d-3)}_{(W)}&=\frac{1}{d-3}\left(-\partial_A W \xbar A^A+W\xbar A_r\right)\,,\\
\mu^{(d-4)}_{(W)}&=-\partial_A W\xbar D^A\Psi^{(d-5)}+(d-4)W\Psi^{(d-5)}\,,\\
\mu^{(d-3)}_{(W)}&=\frac{1}{(d-3)\sqrt{\xbar g}}\Big[ -\partial_A W \Big(\xbar \pi^A+\sqrt{\xbar g}\xbar D^A \Psi^{(d-4)}\Big)+W\Big(\xbar \Pi^r-\sqrt{\xbar g}\Psi^{(d-4)}\Big)\Big]\,.
\end{align}

Therefore, due to $\mathcal L_{\xi,\xi^i}\Omega=0$, the theory is invariant under Poincar\'e transformations. This fact enables us to compute directly the Poincar\'e canonical generators, whose expressions are shown in Section \ref{sec:Section4}.

\section{Transformation laws of the first terms in the asymptotic expansion of the fields} \label{app:AppendixB}
In this appendix we list the transformation laws of the first terms in the asymptotic expansion of the fields under Poincar\'e symmetry. These are obtained by asking for preservation of the asymptotic conditions.
\begin{itemize}
\item Leading orders:
\begin{align}
\delta_{\xi,\xi^{i}}\Phi_{\text{lin}} & =\mathcal{L}_{Y}\Phi_{\text{lin}}+b\Psi_{\text{lin}}\,,\\
\delta_{\xi,\xi^{i}}\xbar \Phi & =\mathcal{L}_{Y}\xbar\Phi+\partial_A W \xbar D^A\Phi_{\text{lin}}+W\Phi_{\text{lin}}+b\xbar \Psi+T\Psi_{\text{lin}},,\\
\delta_{\xi,\xi^{i}}\Psi_{\text{lin}} & =\mathcal{L}_{Y}\Psi_{\text{lin}}+\xbar D_{A}\left(b\xbar D^{A}\Phi_{\text{lin}}\right)+(d-1)b\Phi_{\text{lin}}\,,\\
\delta_{\xi,\xi^{i}}\xbar \Psi & =\mathcal{L}_{Y}\xbar \Psi+\partial_A W\xbar D^A\Psi_{\text{lin}}+\xbar D_{A}\left(b\xbar D^{A}\xbar \Phi\right)+T\left(\xbar \triangle\,\Phi_{\text{lin}}+(d-2)\Phi_{\text{lin}}\right)\,.
\end{align}
\item Core fields:
\begin{align}
\delta_{\xi,\xi^{i}}\xbar A_{r} & =\mathcal{L}_{Y}\xbar A_{r}-(d-4)\partial_A W\xbar D^A\Phi^{(d-5)}+(d-5)(d-4)W\Phi^{(d-5)}\nonumber\\
&\quad+\frac{b}{\sqrt{\xbar g}}\xbar\pi_{r}-(d-4)b\Psi^{(d-4)}-(d-4)T\Psi^{(d-5)}-(d-4)\epsilon^{(d-4)}_{(T,W)}\,,\\
\delta_{\xi,\xi^{i}}\xbar A_{A} & =\mathcal{L}_{Y}\xbar A_{A}+\partial_{A}\left(-(d-5)W\Phi^{(d-5)}+\xbar D^{B}W\partial_{B}\Phi^{(d-5)}\right)\nonumber\\
&\quad+\frac{b}{\sqrt{\xbar g}}\xbar\pi_{A} +\partial_{A}\left(b\Psi^{(d-4)}+T\Psi^{(d-5)}\right)+\partial_A \epsilon^{(d-4)}_{(T,W)}\,,\\
\delta_{\xi,\xi^{i}}\xbar\pi^{r} & =\mathcal{L}_{Y}\xbar\pi^{r}+\sqrt{\xbar g}\,\xbar D_{A}\left[b\left(\xbar D^{A}\xbar A_{r}+(d-4)\xbar A^A\right)\right]\,,\\
\delta_{\xi,\xi^{i}}\xbar\pi^{A} & =\mathcal{L}_{Y}\xbar\pi^{A}-\sqrt{\xbar g}\,\xbar g^{AB}\xbar D^{C}\big(b\xbar F_{BC}\big)+(d-5)\sqrt{\xbar g}\,b\left(\xbar D^{A}\xbar A_{r}+(d-4)\xbar A^A\right)\,,
\end{align}

\item Subleading orders:
\begin{align}
\delta_{\xi,\xi^{i}}A_{r}^{(2)} & =\mathcal{L}_{Y}A_{r}^{(2)}+\xbar D^{A}W\left(\partial_{A}\xbar A_{r}-\xbar A_{A}\right)-(d-3)W\xbar A_{r}\nonumber \\
 & \quad+\frac{b}{\sqrt{\xbar g}}\pi_{(2)}^{r}-(d-3)b\Psi^{(d-3)}+\frac{T}{\sqrt{\xbar g}}\xbar\pi^{r}-(d-3)T\Psi^{(d-4)}-(d-3)\epsilon^{(d-3)}_{(T,W)}\,,\\
\delta_{\xi,\xi^{i}}A_{A}^{(2)} & =\mathcal{L}_{Y}A_{A}^{(2)}+\xbar D_{B}\xbar D_{A}W\xbar A^{B}+\xbar D^{B}W\xbar D_{B}\xbar A_{A}+\partial_{A}W\xbar A_{r} \nonumber\\
&\quad -(d-4)W\xbar A_A+\frac{b}{\sqrt{\xbar g}}\pi_{(2)A}+\frac{T}{\sqrt{\xbar g}}\xbar\pi_{A}+\partial_{A}\left(b\Psi^{(d-3)}+T\Psi^{(d-4)}\right)+\partial_A \epsilon^{(d-3)}_{(T,W)} \,,\\
\delta_{\xi,\xi^{i}}\pi_{(2)}^{r} & =\mathcal{L}_{Y}\pi_{(2)}^{r}+\xbar D_{A}\left(\xbar D^{A}W\xbar\pi^{r}\right)-\partial_{A}W\xbar\pi^{A}\nonumber \\
 & \quad+\sqrt{\xbar g}\,\xbar D_{A}\big[b\big(\xbar D^{A}A_{r}^{(2)}+(d-3)A^{(2)A}\big)\big]+\sqrt{\xbar g}\,\xbar D_{A}\left[T\left(\xbar D^{A}\xbar A_{r}+(d-4)\xbar A^A\right)\big)\right]\,,\\
\delta_{\xi,\xi^{i}}\pi_{(2)}^{A} & =\mathcal{L}_{Y}\pi_{(2)}^{A}+\xbar D_{B}\left(\xbar D^{B}W\xbar\pi^{A}\right)+\xbar D_{B}\xbar D^{A}W\xbar\pi^{B}+\xbar D^{A}W\xbar\pi^{r}-W\xbar\pi^{A} \nonumber\\
&\quad-\sqrt{\xbar g}\,\xbar g^{AB}\xbar D^{C}\big(bF_{BC}^{(2)}\big)+(d-4)\sqrt{\xbar g}\,b\big(\xbar D^{A}A_{r}^{(2)}+(d-3)A^{(2)A}\big)\nonumber \\
 & \quad-\sqrt{\xbar g}\,\xbar g^{AB}\xbar D^{C}\big(T\xbar F_{BC}\big)+(d-4)\sqrt{\xbar g}\,T \left(\xbar D^{A}\xbar A_{r}+(d-4)\xbar A^A\right)\,,
\end{align}
\begin{align}
\delta_{\xi,\xi^{i}}\Psi^{(d-4)} & =\mathcal{L}_{Y}\Psi^{(d-4)}+\partial_{A}W\xbar D^{A}\Psi^{(d-5)}-(d-4)W\Psi^{(d-5)}\nonumber\\
&\quad\xbar D_{A}\left(b\xbar A^{A}\right)+2b\xbar A_{r}+T\left(\xbar\triangle\,\Phi^{(d-5)}-2(d-5)\Phi^{(d-5)}\right)+\mu_{(b,T,W)}^{(d-4)}\,,\\
\delta_{\xi,\xi^{i}}\Psi^{(d-3)} & =\mathcal{L}_{Y}\Psi^{(d-3)}+\partial_{A}W\xbar D^{A}\Psi^{(d-4)}-(d-3)W\Psi^{(d-4)}\nonumber \\
 & \quad+\xbar D_{A}\left(bA^{(2)A}\right)+bA_{r}^{(2)}+T\left(\xbar D_{A}\xbar A^{A}+\xbar A_{r}\right)+\mu_{(b,T,W)}^{(d-3)}\,.
\end{align}
\end{itemize}
The corrective gauge parameters are given by
\begin{align}
\epsilon^{(d-4)}_{(T,W)}&=-T\Psi^{(d-5)}-\partial_A W \xbar D^A \Phi^{(d-5)}+(d-5)W\Phi^{(d-5)}\,,\\
\epsilon^{(d-3)}_{(T,W)}&=\frac{1}{d-3}\left(-T\Psi^{(d-4)}-\partial_A W \xbar A^A+W\xbar A_r\right)\,,\\
\mu^{(d-4)}_{(b,T,W)}&=-\left[2b\xbar A_r+T\left(\xbar \triangle \Phi^{(d-5)}-2(d-5)\Phi^{(d-5)}\right)+\partial_A W\xbar D^A\Psi^{(d-5)}-(d-4)W\Psi^{(d-5)}\right]\,,\\
\mu^{(d-3)}_{(b,T,W)}&=\frac{1}{d-3}\Big[-2(d-2)bA^{(2)}_r-T\xbar D_A \xbar A^A-(2d-5)T\xbar A_r\nonumber\\
&\qquad \qquad \,\,\, -\frac{1}{\sqrt{\xbar g}}\partial_A W \Big(\xbar \pi^A+\sqrt{\xbar g}\xbar D^A \Psi^{(d-4)}\Big)+\frac{W}{\sqrt{\xbar g}}\Big(\xbar \pi^r-(d-5)\sqrt{\xbar g}\Psi^{(d-4)}\Big)\Big]\,.
\end{align}

\end{document}